# Exploring Diffusion Characteristics that Influence Serious Games Adoption Decisions

Katerina Antonopoulou [a] & Nicholas Dacre [a]

[a] University of Warwick Business School, University of Warwick, Coventry, CV4 7AL, UK

**Abstract**

In this paper we discuss the diffusion of serious games and present reasons for why Rogers' traditional approach is limited in this context. We present an alternative overview through the characteristics of relative advantage, compatibility, complexity, trialability, and observability, that reflect on the adoption decision and contributes on the commercialization of serious games.

**Keywords**: Diffusion, Innovation, Virtual Environments, Serious Games.

## Introduction

The growing adoption of serious games and the proliferation of digital technologies which incorporate them, have led to a rapid reshaping of organizations, markets and social infrastructures [1-3]. Companies involved in the development of such innovative products face major changes in implementing a diffusion strategy, since these digital innovations reconfigure the social and technical environment around them [4, 5], which result in further challenges for prospective customers who in turn struggle to understand their potential contribution. Within this context, serious games are considered as a case of digital innovation. For example, Yoo, Henfridsson and Lyytinen state that digital innovations are artifacts constituted by a composition of digital and physical characteristics, with attributes of digital innovation namely including: re-programmability, homogenization and self-reference [6]. Therefore, this paper focuses on serious games, since they exhibit the characteristics of digital innovations in addition to facing challenges regarding their implementation.

## Literature Review

Over the last decade the digital games development community has come to recognize that digital games can be used beyond entertainment purposes [7-9]. These can also be used to train, educate, investigate and advertise [2]. For instance, as users become 'active learners' in a virtual environment that simulates their reality [10], they can be educated in taking appropriate levels of risk or finding solutions in overcoming specific challenges. The advantage is that users receive instant feedback on their actions and as a result can enhance their critical reasoning and decision-making skills [2].

Another area to consider is that employees can be trained in fulfilling their responsibilities with greater efficiency by dealing with common challenges as well as unexpected events [7, 8]. Such as with "Fire Department Training" [11] which was created by the Virtual Worlds Lab at the Georgia Institute of Technology [12]. It is a simulator that enables firefighters to train in different emergency situations in order to develop their skills by combating fires in a virtual environment. Furthermore, a number of games have been developed by the Comparative Media Studies Program at MIT in order to meet varying preparative requirements [13]. One such serious game is 'Biohazard: Hot zone', a game designed "to help emergency first responders deal with toxic spills in public locations" [14]. It offers training for emergency situations that require immediate attention such as a catastrophic incident caused by a chemical leak in a suburban shopping center [2].

Despite the growing implementation of serious games, for some authors serious games have failed to attain the expected levels of success and profitability despite the increasing technical capabilities and realistic settings [15, 16]. Furthermore, the benefit that developers of serious games offer to potential users, are limited by client's constraints to realize the usefulness of these digital innovations [15, 17]. In terms of digital innovation attributes, re-programmability offers a distinction between the operational logic and the realized use of the object by consumers [6]. The operational logic of a serious game can be the graphical and resource management architecture which aim to represent a strategy that specifies how a process operates through a system with specific participants and behaviors [18]. Whilst the realized use of a serious game, by the actors, is the actual game or platform which includes the various stages of engagement and scenario architecture.

Compatibility is the combination of digital devices, with digitized data, and offers a wide range of services, products and information [6]. For example, the distribution of knowledge for a specific incident from a digital device is feasible because the 'information' is formed into 'digital information' that is compatible with a digital device. Thus, comparable digital devices are able to reproduce or process the digitized information. The aforementioned characteristics of serious games as a case of digital innovations can reconfigure the traditional diffusion characteristics. As such, the purpose of this paper is to provide an overview of the diffusion of serious games and present reasons for which traditional innovation characteristics as developed by Rogers [19], do not provide a consistent rationale [20] in support of serious games adoption decision. Therefore, in the following section we draw on Roger's [19] innovation diffusion model to provide an overview of these salient points.

## Conceptual Approach

The adoption of innovations has been an important focus for researchers since the seventies, with Rogers' [19] diffusion model being one of the most widely used to date. Rogers defines diffusion as "the process in which an innovation is communicated through certain channels over time among the members of a social system" [19].

Diffusion is referred to as distinctive form of communication, where the content is premised on the distribution of new ideas. This process requires sharing information between actors towards a convergent understanding, however as the diffusion of innovation elaborates on novel concepts, the practice can also further divide shared meaning between individuals.

Diffusion of innovation theory focuses on the perceived meanings of certain events by the end users [21] and includes five distinct characteristics which have a direct impact on the adoption process of innovations. These characteristics are specified as follows: relative advantage, compatibility, complexity, trialability, and observability [19]. Through these, Rogers' diffusion of innovation model whilst focusing specifically on the decision process provides an approach to explore serious games through the perceived attributes of innovation adoption. Therefore, this paper draws on Rogers model and focuses specifically on the diffusion of serious games and the direct impact on their adoption as a decision of "full use of an innovation as the best course of action available" [19].

Prior research within the Information System (IS) discipline has sought to understand the existing characteristics of Rogers' model, and develop these further through a combination of removing specific characteristics and integrating other elements through validation [22-26]. However, the theoretical approach to diffusion characteristics in relation to serious games, as a case of digital innovation, is currently limited in academic research. This presents an opportunity to explore the attributes of diffusion for serious games, by discussing reasons which may be identified as lacking in the traditional set of characteristics, and thus do not provide a consistent rationale [20] in support of serious games adoption.

## Discussion

The attributes of diffusion as delineated by Rogers [19], stipulate the five characteristics within the decision process of relative advantage, compatibility, complexity, trialability, and observability. These particular characteristics are discussed in the following literature, towards building a broader understanding of innovations adoptions within the context of serious games.

Relative advantage is "the degree to which an innovation is perceived as being better than the idea it supersedes" [19]. Although serious games do not generally replace traditional training methods, by applying game thinking and transforming knowledge into digital knowledge, serious games can increase the quality of training and decision making to a higher level [27-29]. However potential users of serious games may attempt to recognize the "intensity of the reward or punishment resulting from adoption of an innovation" [30], as to whether serious games can improve processes and increase their returns on investment, although it is noted that an improvement to business processes can reduce costs by as much as 90% [31].

Rogers states that relative advantage can be expressed as economic profitability or social status and has a direct impact on any innovations' rate of adoption [19]. However, relative advantage is intertwined with economic profitability, a low initial cost, a reduction in discomfort, a savings in time and effort, and the immediacy of the reward [19]. As Dougherty [32] argues, potential actors may not necessarily preconceive requirements and the set of attributes of innovations, as these may diverge prior and after the use of a digitized product. Individuals and firms are surrounded by vast amounts of data which they strain to interpret and apply towards ensuring optimum information-based decisions [33]. Thus, potential adopters of serious games are cognizant of the demand in discerning and analyzing data, towards extant solutions which can be adjusted to different circumstances. Conversely, the potential of serious games in facilitating such processes has not been as widely affirmed as anticipated [15, 17].

Universal adoption by potential customers have yet to recognize through innovations diffusion, that decision making and training with serious games can enhance the assimilation of novel skills, and that the digital nature of these artifacts permits their customization. For instance, following initial investment additional scenarios or stages can be applied to serious games towards training employees in different situations [7, 8]. Equally this provides the opportunity to leverage decision making skills in an innocuous environment where training is founded on volatile and dangerous contexts [28, 34]. Furthermore, the financial commitment for training may be reduced following the initial outlay, since fewer employees are required in either training new members or in actually delivering the content.

Complexity is "the degree to which an innovation is perceived as relatively difficult to understand and use"[19]. Serious games are designed to enhance skills such as decision making and cooperative engagement by providing high quality training for individuals [27]. Through this process actors can increase their cognate responses by adapting to novel challenges [11]. Although having previously outlined these potential benefits, serious games developers must overcome innovations diffusion challenges in convincing clients of the usefulness and convenience of serious games [15]. Moreover, Rogers states that the perceived complexity of an innovation is negatively related to its adoption [19]. It would thus appear, that organizations are contrived and reluctant in investing in radical innovation, as it may be challenging to accumulate new capabilities and relinquish prior existing ones [35]. Furthermore, serious games are considered as social artifacts that are implemented in a social system where there exists societal resistance [36].

In this context, potential customers struggle to distinguish that serious games offer a virtual environment that simulates reality, but also embody game characteristics such as 'fun' and 'satisfaction' [37]. As a result, given the variety of skills and varying different knowledge which may be gained through serious games, customers may perceive this technology inappropriate and complex as a non-entertainment platform [17]. This raises challenges in the development of serious games in persuading end-users that these provide a viable training and skills development platform [38].

Compatibility "is the degree to which an innovation is perceived as consistent with existing values, past experiences, and needs of potential adopters" [19]. As one of the five characteristics of innovations, compatibility is a compelling diffusion factor concerning the perceived values of serious games. As previously stated, apart from technological artifacts, serious games are also social artifacts that redefine the traditional theories concerning the evaluation of such technologies [36]. The recognized present values of traditional methods of training and education incorporate different theorizations of value. These compete through the perceived values of serious games and how they interplay with pre-existing values. Novel approaches form a sensitive issue that can affect the adoption of serious games. Thus it is noteworthy to consider that several studies have been conducted for either the economic or social value of innovations. Of which, the breadth of perspectives in the literature, reveal that controversies between socioeconomic values, challenge many of the traditional views of innovation in an attempt to provide a consistent rationale [6, 39, 40].

As serious games create high uncertainty through novel technology, these challenge established enterprise philosophies and the process in which they operate [1-3]. They create a distinct market as they influence the internal structures of the firm, and equally external relations with other firms. The trajectory of existing experiences and requirements, play a vital role for developers during the production process of serious games. Furthermore, developers must demonstrate that newly acquired skills and knowledge through serious games can actively be used beyond virtual settings and applied to the benefit of the organization. Within this context the attribute of trialability is examined.

Trialability is described as the "degree to which an innovation may be experimented with on a limited basis" [19]. Trialability represents the opportunity by which an actor is able to engage with a novel product, service or entity standing as an innovation. The greater the factors in gaining access to innovations and having the resources to test these, the greater the likelihood of adopting these. By applying this characteristic to serious games, providing a 'testbed' approach where clients are able to directly engage through the development stages, would provide a greater understanding and input into the platform. This would enable potential customers to better appreciate the integration between learning and playing, and navigate through a virtual environment with actual consequences and experiences. Thus, serious games can further maintain relevance and authenticity, but also encourage extrinsic and intrinsic motivation in user participation. Although this may not be attributable to all scenarios, for instance serious games developed for military and emergency services [2], may prove more challenging in their testing approach.

Serious games developers face various challenges in clearly delineating the power these can leverage in terms of training and education away from misinterpretations of vacuous play [41, 42], towards cultivating coordinated and cooperative skills that are needed for specific and highly synchronized settings. Therefore, when trialability is not an appropriate approach, then other processes undertake the involvement of potential users in the development of a serious game. When this process occurs, then it is easier to bridge the perceptions of developers and potential customers, and the perceived usefulness of the game, however then the concept of diffusion reconfigures again since developers will seek methods to engage potential customers, as the challenge of diffusion is an antecedent to the development of the artifact.

Finally, observability is the degree "an innovation is visible to others" [19]. Through this characteristic when the impact of a digital innovation offers a greater degree of visibility, this in turn elicits a greater positive, or albeit negative, reaction from actors. Additionally, serious games consist of presently observable artifacts as similarly experienced in a game context, although it may be challenging for developers to portray to end-users how these artifacts reconfigure the traditional practices and uses through the medium of digitalization. Although degrees of observability may be limited in scope, the advantages of serious games are easily perceived through the processes of firms once implemented. Additionally, a serious game can be customized to the client's requirements, when there is a setting or part of the architecture needing change.

## Conclusion and Further Research

Rogers' diffusion characteristics have been proved to be a very useful theoretical approach that has a direct impact on IS implementation. Several models have been developed drawing on trialability, relative advantage, complexity, compatibility and observability, and have been tested empirically to prove the reliability and robustness of the models. In the current research, we explore the diffusion of serious games and present reasons for which the aforementioned characteristics do not provide a consistent rationale in support of serious games adoption. Further to this research, attempts will be made to mirror several contributions on the serious game side. An empirical analysis of these factors and an integration of further factors such as re-programmability, self-reference and homogenization will be tested towards developing a model that will yield statistically reliable results. Our intended contribution is twofold; elucidate our understanding of the adoption decision of serious games, and by drawing on such an understanding, outline the implications for the commercialization of serious games. As such, a reliable adoption model will enable further research to how business plans and business models reshape and trigger new practices in order to assess the market for serious games.